\documentclass[11pt]{article}
\usepackage{latexsym}
\usepackage{amssymb}
\usepackage{epsf}
\usepackage{amsmath}
\usepackage{bm}
\usepackage{graphicx}

\begin{document}
\pagestyle{empty}
\renewcommand{\thefootnote}{\fnsymbol{footnote}}

\vspace*{1cm}

\begin{center}

{\bf \LARGE Holographic Dual of Quantum Spin Chain \\
as Chern-Simons-Scalar Theory} 
\\

\vspace*{2cm}
{\large 
Naoto Yokoi$^{1, 2}$\footnote{yokoi@g.ecc.u-tokyo.ac.jp}, Yasuyuki Oikawa$^{3}$\footnote{yasuyuki.oikawa.e2@tohoku.ac.jp}  
and Eiji Saitoh$^{1, 2, 3}$\footnote{eizi@ap.t.u-tokyo.ac.jp}} \\
\vspace*{1.5cm}

{\it $^{1}$Department of Applied Physics, The University of Tokyo, Tokyo 113-8656, Japan \\
$^{2}$Institute for AI and Beyond, University of Tokyo, Tokyo 113-8656, Japan \\
$^{3}$WPI Advanced Institute for Materials Research, Tohoku University, 
Sendai 980-8577 Japan} 
\end{center}

\vspace*{3cm}

\begin{abstract}
{\normalsize 
We construct a holographic dual theory of one-dimensional anisotropic Heisenberg spin chain, 
which includes two Chern-Simons gauge fields and a charged scalar field.   
Thermodynamic quantities of the spin chain at low temperatures, which are exactly calculated from the integrability,  
are completely reproduced by the dual theory on three-dimensional black hole backgrounds and 
the exact matching of the parameters between the dual theory and the spin chain is obtained.     
The holographic dual theory provides a new theoretical framework to analyze the quantum spin chain 
and one-dimensional quantum many-body systems.  
} 
\end{abstract} 

\newpage
\baselineskip=18pt
\setcounter{page}{2}
\pagestyle{plain}
\baselineskip=18pt

\renewcommand{\thefootnote}{\arabic{footnote}}
\setcounter{footnote}{0}

\section{Introduction}
The quantum spin chain has been serving a role as an important touchstone for 
the study on dynamics of quantum many-body systems, such as quantum entanglement \cite{BayatEntanglement} and 
quantum phase transition \cite{SachdevQPT}.  
In the analysis of the quantum spin chain, several theoretical frameworks have been developed:  
the Bethe ansatz, bosonization, and (conformal) field theory techniques \cite{FranchiniBethe, Giamarchi, Fradkin}.  
Thanks to these techniques, the quantum spin chain has been established as a quantum integrable model, 
and allowed to have some exact results from the analytical calculations.      

Recently, for the nonperturbative analysis of quantum many-body systems, 
yet another framework has been emerged: the holographic duality \cite{ZaanenHolography, Hartnoll:2018xxg}. 
The holographic duality, which is also known as the AdS/CFT correspondence \cite{Maldacena:1997re}, is 
a correspondence between quantum many-body systems (or quantum field theories) in $d$-dimensional space and 
gravitational theories in $(d+2)$-dimensional curved spacetime. 
So far, the holographic duality has been applied to various condensed matter systems,  
such as superconductors \cite{Hartnoll:2008vx, Hartnoll:2008kx, Horowitz:2010gk} and magnetic materials  
\cite{Iqbal:2010eh, Cai:2014oca, Cai:2015jta, Yokoi:2015qba, Yokoi:2019waj, Yokoi:HolographicReview}. 
Although those analyses based on the holographic duality qualitatively reproduce the critical behavior 
of temperature and external field dependences,  
few examples can successfully reproduce the physical quantities 
at the quantitative level in the applications to condensed matters. 
The quantum spin chain has the exact results based 
on various integrable techniques, and thus we can compare the physical quantities calculated from 
the holographic dual theory with such exact results quantitatively including numerical factors.\footnote{Integrability 
has been also discussed in the AdS/CFT correspondence from the string theory perspectives. See \cite{Beisert:2010jr} 
for a review and references therein.}   

In this Letter, we construct the holographic dual theory of one-dimensional antiferromagnetic Heisenberg spin chain 
($S=1/2$) with anisotropy, which is three-dimensional abelian Chern-Simons gauge theory 
coupled with a charged scalar field,  and show that the dual theory on three-dimensional black holes  
can analytically and quantitatively reproduce the physical quantities of the spin chain in the low-energy regions.    
In the course of this study, the matching between the parameters in the both theories are 
determined including numerical factors.    
From the three-dimensional perspectives, the dual gauge theory can lead   
to a new theoretical framework to analyze the quantum spin chain and also one-dimensional quantum many-body systems.

\section{Spin chain at low temperatures as Chern-Simons theory}
The antiferromagnetic quantum spin chain ($S=1/2$) with the anisotropy along the $S^{z}$ direction (XXZ spin chain) 
is defined by the Hamiltonian
\begin{equation}
\mathcal{H} = J \sum_{i=1}^{N} \left( S^{x}_{i}S^{x}_{i+1} + S^{y}_{i}S^{y}_{i+1} + \delta\, S^{z}_{i}S^{z}_{i+1} \right) ,
\label{eq: spin-chain Hamiltonian}
\end{equation}
where $J>0$ is the exchange coupling constant, and $\delta = - \cos(\pi\beta^2)$ with $0\le\beta^2\le1$ is 
the anisotropy parameter. 
Using the Jordan-Wigner transformation and the bosonization technique, 
the dynamics of the XXZ spin chain 
can be described by the Gaussian model of a compactified boson $\Phi$ 
with the radius $R(\beta) = \beta/\sqrt{2\pi}$, which is a $(1+1)$-dimensional conformal field theory (CFT) 
with the central charge $c=1$, in the continuum limit 
at low temperatures \cite{Giamarchi, Fradkin}. 
The Gaussian model possesses the chiral $U(1)_{L} \times U(1)_{R}$ symmetry whose currents are given 
by $J_{z}^{(L)} \propto \partial_{z} \Phi$ and $J_{\bar{z}}^{(R)} \propto \partial_{\bar{z}} \Phi$ respectively. The both chiral currents 
suffer from the chiral anomaly, and only the vector-like combination of the currents satisfies the conservation law, 
which corresponds to the total spin conservation, $S^{z} = \sum_{i=1}^{N} S_{i}^{z}$. 
From the calculation based on the CFT technique and the Bethe ansatz method,  
the susceptibility for the spin density ($s^{z} = S^{z}/N$) in the low temperature limit ($T \rightarrow 0$) 
has been exactly obtained to be 
\begin{eqnarray}
\chi_{0} = \frac{\partial \langle s^{z} \rangle}{\partial H} = \frac{1}{2 \pi\,\beta^{2}\,v_{s}} 
\qquad \textrm{with} \qquad \beta 
= \sqrt{1 - \cos^{-1} \delta/\pi} ,
\label{eq: Affleck susceptibility}
\end{eqnarray}  
where $H$ is the external magnetic field and $v_{s}$ is 
the spin-wave velocity \cite{Affleck:1994}. In the following discussion, 
we take the unit of  $v_{s} = 1$ for simplicity. 

The holographic dual theory of the XXZ quantum spin chain 
in the low temperature limit is given by the $U(1)_{L} \times U(1)_{R}$ 
Chern-Simons (CS) gauge theory on 
a three-dimensional manifold $M$,\footnote{This type of the CS gauge theory has been also discussed in 
the AdS$_{3}$/CFT$_{2}$ correspondence 
with the string theory constructions \cite{Kraus:2006nb, Kraus:2006wn, Jensen:2010em}} whose action is given by
\begin{eqnarray}
S_{\textrm{CS}}[A, B] = - \frac{k}{4 \pi} \int_{M}\!\!\! d^{3}x\, \epsilon^{\mu\nu\rho} 
\left( A_{\mu} \partial_{\nu} A_{\rho} - B_{\mu} \partial_{\nu} B_{\rho} \right) .
\label{eq: CS action}
\end{eqnarray}
Here, $A_{\mu}$ and $B_{\mu}$ are the gauge fields of $U(1)_{L}$ and $U(1)_{R}$ respectively, 
$k$ is a positive constant, and $\epsilon^{\mu\nu\rho}$ is the (normalized) third-rank totally antisymmetric tensor. Note that this action does not depend on the metric of three-dimensional base space, and thus the CS theory is topologically invariant.  

The holographic duality assumes the negatively curved spacetime with the (conformal) boundary, so-called asymptotically 
Anti-de Sitter (AdS) spacetime, as the background geometry.  
Here, for concreteness, we consider the base space as the Euclidean three-dimensional AdS spacetime 
(AdS$_{3}$), whose metric is given by the Poincar\'e metric\,:
\begin{eqnarray}
ds^{2} = \frac{\ell^2}{y^2} \left(dy^2 + dz d\bar{z}\right) ,
\label{eq: Poincare metric}
\end{eqnarray}
where $z = x + i\,t$ and $\bar{z} = x - i\,t$, and we take $\epsilon^{y z \bar{z}} = 1$. 
Note that the background geometries are given by 
the solutions of Einstein equation 
with a negative cosmological constant, $\Lambda = - 1/\ell^2$, 
and we consider the probe approximation ignoring further dynamics of gravity 
in the following discussion.     

The CS action (\ref{eq: CS action}) is gauge invariant, and we take $A_{y} = B_{y} = 0$ 
as the gauge fixing condition. 
Using this gauge fixing and the parametrization (\ref{eq: Poincare metric}), 
the bulk equations of motion are given by the flatness conditions:
\begin{eqnarray}
&&\partial_{y} A_{z} = \partial_{y} B_{z} = 0, \qquad F^{(A)}_{z \bar{z}} = F^{(B)}_{z \bar{z}} = 0 .
\label{eq: CS eq of motion}
\end{eqnarray} 
Here, in order to cancel the boundary terms, we include the boundary action 
as a counter term, following Ref.\,\cite{Keranen:2014ava}:  
\begin{eqnarray}
S_{\textrm{bd}} = - \frac{k}{4 \pi}  \int_{y\textrm{=}0}\!\!\! d^{2}z\,\left( A_{z} A_{\bar{z}} +  
B_{\bar{z}} B_{z} - 2 A_{\bar{z}} B_{z} \right) .
\label{eq: correct boundary term}
\end{eqnarray}
with the Dirichlet boundary condition, $\delta A_{\bar{z}} = \delta B_{z} = 0$ at $y = 0$.   
In the context of the holographic duality, the boundary action and the boundary condition 
imply that $A_{\bar{z}}$ and $B_z$ correspond to the external sources for the chiral current operators  
$J_{z}^{(L)}$ (dual to $A_z$) and $J_{\bar{z}}^{(R)}$ (dual to $B_{\bar{z}}$) in the 
boundary CFT.   

In the holographic duality, 
the Gubser-Klebanov-Polyakov-Witten (GKPW) relation 
\cite{Gubser:1998bc, Witten:1998qj} claims the equivalence of 
the partition functions of the quantum many-body system and 
the holographic dual theory: $Z_{\textrm{qms}} = Z_{\textrm{hol}} 
\simeq e^{- S_{\textrm{hol}}}$, where the last relation is the semiclassical approximation 
in the dual theory.  
From this relation, we can obtain the expectation values of 
the current operators in the boundary CFT :
\begin{eqnarray}
\langle J^{(L)}_{z} \rangle_{A_{\bar{z}}} &=& - \left.\frac{\delta \left(S_{\textrm{CS}} + 
S_{\textrm{bd}}\right)}{\delta A_{\bar{z}}}\right|_{\textrm{on-shell}} = \left[\frac{k}{2 \pi} A_{z} -  
\frac{k}{2 \pi} \bigl. B_{z} \right]_{y=0}\, , \nonumber \\
\langle J^{(R)}_{\bar{z}} \rangle_{B_{z}} &=& - \left.\frac{\delta \left(S_{\textrm{CS}} + 
S_{\textrm{bd}}\right)}{\delta B_{z}}\right|_{\textrm{on-shell}}  = \left[\frac{k}{2 \pi} \bigl. B_{\bar{z}} 
- \frac{k}{2 \pi} \bigl. A_{\bar{z}} \right]_{y=0}\, .
\label{eq: expectation value of chiral current}
\end{eqnarray}  
where the equations of motion (\ref{eq: CS eq of motion}) have been used for the on-shell evaluation.   
From the conformal invariance, the left-current $J^{(L)}_{z}$ should be a holomorphic function 
and the right-current $J^{(R)}_{\bar{z}}$ 
should be an anti-holomorphic function  
and thus these should satisfy the conditions,
$\partial_{\bar{z}} J^{(L)}_{z} = 0$ and $\partial_{z} J^{(R)}_{\bar{z}} = 0$ on the boundary.   
However, the holographic expectation values (\ref{eq: expectation value of chiral current}) imply  
\begin{eqnarray}
\partial_{\bar{z}} \langle J^{(L)}_{z} \rangle =  
\frac{k}{2 \pi} \left(\partial_{z} A_{\bar{z}} - \partial_{\bar{z}} B_{z}\right), \qquad 
\partial_{z} \langle J^{(R)}_{\bar{z}} \rangle =  
\frac{k}{2 \pi} \left(\partial_{\bar{z}} B_{z} - \partial_{z} A_{\bar{z}}\right) ,
\end{eqnarray}
where the flatness conditions of the gauge fields have been used.    
This relation corresponds to the chiral anomaly in the boundary CFT (the Gaussian model), which spoils the $U(1)_{L} \times U(1)_{R}$ 
invariance under the nontrivial background gauge fields, $A_{\bar{z}}$ 
and $B_{z}$ \cite{Kraus:2006nb, Kraus:2006wn, Jensen:2010em}. 
However, it is important that we have the conserved vector-like Noether current, 
$J = J_{z}^{(L)} dz + J_{\bar{z}}^{(R)} d\bar{z}$, on the boundary:\,\footnote{Our convention for 
the two-dimensional current is $J_{z} = \frac{1}{2}\left(J_{x}-i J_{t}\right),~ J_{\bar{z}} = 
\frac{1}{2}\left(J_{x}+i J_{t}\right)$.} 
\begin{eqnarray}
d\, \left(* J\right) = \partial_{\bar{z}} \langle J^{(L)}_{z} \rangle + \partial_{z} \langle J^{(R)}_{\bar{z}} \rangle = 0 . 
\label{eq: current conservation}
\end{eqnarray}
This is consistent with the total spin conservation in the XXZ spin chain.  
Note that the conservation is attributed to the last term in the boundary action (\ref{eq: correct boundary term}), 
which vanishes under the variation with the boundary condition \cite{Keranen:2014ava}.  

The conserved charge density, which corresponds to the spin density $s^{z}$, 
is given by $(J^{(L)}_{z} - J^{(R)}_{\bar{z}})/2 \propto J_{t}/2$ and, accordingly, the conjugate external source 
for the spin density corresponds to the combination of gauge fields, 
$H = A_{\bar{z}} - B_{z}$.  
For the regularity of general bulk geometries, we assume 
the vanishing temporal components of gauge fields ($A_{t}=B_{t}=0$) and 
only consider the non-vanishing spatial components of gauge fields 
\cite{Kraus:2006nb, Kraus:2006wn}.\footnote{As discussed below, the backgrounds of AdS$_{3}$ 
and BTZ black holes correspond to the boundary CFT at zero and finite temperature, respectively. 
The bulk gauge fields need to be consistent 
in the both backgrounds so that the thermal effects can be continuously incorporated in the boundary CFT.}    
Then, the external source for the spin density, \textit{i.e.}, the external magnetic field, can be identified as 
$H = \frac{1}{2}\left(A_{x} - B_{x}\right)$. With this identification, the spin density can be expressed 
as $\langle s^{z} \rangle = \frac{1}{2}\left(\langle J_{z}^{(L)} \rangle - \langle J_{\bar{z}}^{(R)} \rangle\right) = \frac{k}{2 \pi} H$,  
and the spin susceptibility in the low temperature limit is given by
\begin{eqnarray}
\chi_{0} = \frac{\partial \langle s^{z} \rangle}{\partial H} = \frac{k}{2 \pi} ,
\label{eq: zero temperature susceptibility}
\end{eqnarray}
and the susceptibility implies that the free energy density of the XXZ spin chain is 
\begin{eqnarray}
\frac{F_{XXZ}^{(0)}}{\ell} = \frac{k}{4 \pi} H^{2} . 
\label{eq: zero temperature free energy}
\end{eqnarray}
Comparing the holographic result with the exact one (\ref{eq: Affleck susceptibility}),  
we have the parameter matching between the dual CS gauge theory and the quantum spin chain: 
\begin{eqnarray}
k = \frac{1}{\beta^{2}} .
\end{eqnarray}

Next, we calculate another thermodynamic quantity, the specific heat in the low temperature limit 
without the external field.  
In the holographic duality, finite-temperature effects of quantum many-body systems can be included  
by introducing the black hole background in the holographic dual theory.  
In the three-dimensional gravity with a negative cosmological constant, 
there exist the black hole solutions,  
so-called BTZ black holes \cite{Banados:1992wn}, which have the asymptotically AdS geometry at $r\sim\infty$.  
In what follows, for simplicity, we consider the non-rotating (Euclidean) BTZ black holes, whose metric is given by  
\begin{eqnarray}
ds^2 = \left(\frac{r^{2}-r_{+}^{2}}{\ell^{2}}\right) dt^2 + \left(\frac{\ell^{2}}{r^{2}-r_{+}^{2}}\right) dr^2 + r^{2} d\theta^2 .
\label{eq: BTZ metric}
\end{eqnarray}  
Here, $r_{+}$ is the radius of the horizon given by $r_{+} = \sqrt{8 G_{N} M}\, \ell$, where $G_{N}$ is three-dimensional 
Newton constant, and $M$ is mass of the black holes. Note that the BTZ black holes have locally same geometry as 
AdS$_{3}$ and the metric (\ref{eq: BTZ metric}) can be 
transformed to the AdS$_{3}$ metric (\ref{eq: Poincare metric}) by the coordinate transformations,
\begin{eqnarray}
y = \left(\frac{r_{+}}{r}\right) e^{\frac{r_{+}}{\ell} \theta} , \qquad z = \sqrt{1 - \frac{r_{+}^{2}}{r^{2}}}~ e^{\frac{r_{+}}{\ell} 
\left(\theta + i \frac{t}{\ell}\right)} .
\end{eqnarray}
The Hawking temperature of the BTZ black hole is given by 
$T_{BH} = \left(\frac{1}{2 \pi \ell}\right) \left(\frac{r_{+}}{\ell}\right)$, which is identified as 
the temperature of the quantum many-body system. 
For convenience, we introduce another coordinate, $u = \frac{\ell^2}{r}$ (and $u_{+} = \frac{\ell^2}{r_{+}}$), 
which leads to 
\begin{eqnarray}
ds^{2} = \ell^2 \left(\frac{f(u)\,dt^{2} + f(u)^{-1} dy^{2} + \ell^{2} d\theta^2}{u^{2}} \right) \quad 
\textrm{with} \quad f(u) = 1 - \frac{u^2}{u_{+}^{2}} . 
\label{eq: BTZ metric in u}
\end{eqnarray}

In the CS gauge theory (\ref{eq: CS action}) with the boundary term (\ref{eq: correct boundary term}),  
the classical on-shell action vanishes without the external field ($A_{x}=B_{x}=0$) and 
the classical partition function is trivial.
In the literature \cite{Porrati:2019knx, Afkhami-Jeddi:2020ezh, Maloney:2020nni}, the quantum one-loop (and exact) partition 
function on the BTZ black hole (\ref{eq: BTZ metric}) is obtained 
by the explicit path-integral calculation of the CS gauge theory: 
\begin{eqnarray}
Z_{CS}(\tau) = \frac{1}{|\eta(\tau)|^{2}} = |q|^{- 1/12}\,\prod_{n=1}^{\infty} \frac{1}{\left|\left(1 - q^{n}\right)\right|^{2}}  
\qquad \left(q = e^{2 \pi i \tau}\right),
\end{eqnarray}   
where $\eta(\tau)$ is the Dedekind's $\eta$-function. 
$\tau$ is the moduli parameter of the boundary torus at $r \rightarrow \infty$, which is given by $\tau = i\,\ell\,T_{BH}$ 
for the BTZ black holes. At low (but finite) temperatures, the partition function can be well approximated by  
$Z_{CS} \simeq \exp\left(\pi\,\ell\,T_{BH}/6\right)$ for $T_{BH} \gtrsim 1/\ell$, and thus the free energy density  
is given by\footnote{$T_{BH} \gtrsim 1/\ell$ indicates $r_{+} \gtrsim 2 \pi \ell$, which guarantees that 
the BTZ black holes are large black holes in AdS$_{3}$ and stable with respect to the Hawking-Page phase transition.} 
\begin{eqnarray}
\frac{F_{CS}}{\ell} = - \frac{T_{BH}}{\ell} \log Z_{CS} \simeq - \frac{\pi}{6} T_{BH}^{2} .
\label{eq: finite size scaling free energy}
\end{eqnarray}
Using the GKPW relation,  
the free energy $F_{CS}$ can be identified with that of the spin chain and we obtain
the specific heat $c_{v} \simeq \frac{\pi}{3} T_{BH}$, which is consistent with 
the exact result based on the finite-size scaling method in CFT  
and the Bethe ansatz method \cite{Bloete:1986qm, Affleck:1986bv}.
In the following discussion, we set the curvature radius of AdS$_{3}$ 
to be unity, $\ell = 1$, for simplicity.

\section{Finite Temperature Correction from Bulk Scalar Field}
We have seen, so far, the conformally invariant effective description of the XXZ spin chain 
in the low temperature limit. 
Here, we consider an irrelevant scalar perturbation of the CFT of the XXZ spin chain  
within the holographic dual theory.   
In this Letter, we consider the scalar perturbation of the 
Sine-Gordon (SG) type, 
\begin{eqnarray}
S_{\textrm{pert}} =  \frac{1}{2}\int\!d^{2}x\, \left( \lambda(x)\, {\cal O}(x) + h. c.\right ) 
= \lambda \int\!d^{2}x~ \cos \left( \Phi(x)/\beta \right) , 
\label{eq: sine-gordon perturbation}
\end{eqnarray}  
where a real constant external source (\textit{i.e.} coupling constant) $\lambda$ is assumed,  
and $\Phi(x)$ is the compactified boson in the CFT description of the spin chain \cite{Lukyanov:1998}. 
The operator ${\cal  O}(x) = e^{i \Phi(x)/\beta}$ has the scaling dimension $d = 2/\beta^{2}$ and the charge 
$(q_{L}, q_{R}) = (1/\beta, -1/\beta)$ under the $U(1)_{L}\times U(1)_{R}$ chiral symmetry.  
The correlation function is normalized to be $\left\langle {\cal O}(x) {\cal O}^{\dagger}(y) \right\rangle = 
\left\langle {\cal O}^{\dagger}(x) {\cal O}(y) \right\rangle = 1/|x - y|^{2 d}$. 
The SG perturbation actually originates from the term proportional to anisotropy ($\sim S^{z}_{i} S_{i+1}^{z}$) 
in the spin-chain Hamiltonian (\ref{eq: spin-chain Hamiltonian}).   

In the context of the holographic duality, the irrelevant scalar perturbations can be described by 
introducing massive bulk scalar fields in the dual theory.     
Thus, we introduce a complex scalar field with the mass $m$ 
in the dual theory, whose action is given by
\begin{eqnarray}
S_{\textrm{sc}} = \int\!d^{3}x\sqrt{g} \left[g^{\mu\nu} (D_{\mu} \phi^{*}) (D_{\nu}\phi) + m^2 \phi^{*} \phi \right] ,
\label{eq: bulk scalar action}
\end{eqnarray}
where the complex scalar has the charge $+\, q$ for $A_{\mu}$ and $-\, q$ for $B_{\mu}$ respectively, and the gauge covariant derivative 
is defined as $D_{\mu} \phi = \partial_{\mu} \phi - i q\,(A_{\mu}-B_{\mu})\,\phi$.  
The resulting equation of motion of $\phi$ from the action (\ref{eq: bulk scalar action}) becomes 
\begin{eqnarray}
\frac{1}{\sqrt{g}} D_{\mu} \left( \sqrt{g} g^{\mu\nu} D_{\nu} \phi \right) - m^2 \phi = 0 .
\label{eq: scalar equation} 
\end{eqnarray}
In the holographic dictionary, the bulk scalar field with the mass $m$ corresponds to the scalar operator 
in the boundary CFT with the scaling dimension, $d = \dim\, {\cal O} = 1 + \sqrt{1 + m^2}$. 
The solutions to the equation of motion have the asymptotic 
expansion near the boundary 
($u\sim 0$):
\begin{eqnarray}
\phi_{\textrm{sol}} \simeq \phi_{0}(x)\, u^{1 - d} + \psi_{0}(x)\, u^{d} + \cdots ,
\end{eqnarray}   
where $x = (t, \theta)$ is boundary coordinates and the dots represent the higher order terms. By means of 
the GKPW relation, the coefficient $\phi_{0}$ gives 
the external source with the renormalization factor, $\lambda/(d-1)$, and $\psi_{0}$ gives the expectation value, 
$\langle{\cal O}(x)\rangle$, under the perturbation (\ref{eq: sine-gordon perturbation}) \cite{Freedman:1998tz, Klebanov:1999tb}.  
Along with the bulk scalar field, we consider finite-temperature corrections to the free energy of the XXZ spin chain  
from the scalar perturbation (\ref{eq: sine-gordon perturbation}) in the holographic dual theory. 

In the perturbed case, the GKPW relation leads 
to the correspondence between the on-shell classical action $S_{\textrm{hol}}$ 
of the CS theory with the scalar field   
and the free energy $F_{\textrm{XXZ}}$ of the XXZ spin chain:  
\begin{eqnarray}
Z_{\textrm{hol}}\left[A, B, \phi\right] = Z_{\textrm{XXZ}}(T, H)~ \Longrightarrow~  
S_{\textrm{hol}}\left[A_{\textrm{sol}}, B_{\textrm{sol}}, \phi_{\textrm{sol}}\right] \simeq F_{\textrm{XXZ}}(T, H)/T .  
\end{eqnarray} 
Here, the on-shell action is evaluated based upon the solutions 
of (classical) equations of motion of the bulk fields. 
In order to obtain the free energy of the XXZ spin chain including finite temperature corrections,
we should have the solution to the equation of motion of the complex scalar field 
on the BTZ black hole (\ref{eq: BTZ metric in u}).

We assume the product form of the solution, $\phi(u, x) = \phi(u)\, \phi_{0}(x)$, and 
further restrict $\phi_{0}$ to a real constant which corresponds to the coupling constant $\lambda$ of 
the SG perturbation.\footnote{We also assume a sufficiently small value of 
$\phi_{0}$ which guarantees the probe approximation ignoring the gravitational backreaction.}  
Using the correspondence between the external magnetic field and the CS gauge fields, 
$H = \frac{1}{2}\left(A_{\theta}-B_{\theta}\right)$, 
the radial equation of motion of the scalar field $\phi(u)$ is explicitly given by  
\begin{eqnarray}
u^2 f(u)\, \partial_{u}^{2} \phi - u \left(1 + \frac{u^{2}}{u_{+}^{2}}\right) \partial_{u} \phi - 
\left(4\,h^2\,u^2 + m^{2}\right) \phi = 0 , 
\end{eqnarray}
where $h = q\,H$ with the charge $q$ of the scalar field and 
the uniform magnetic field ($H=\textrm{const.}$) has been assumed 
for simplicity.   
  
The change of variable, $\zeta = 1 - \frac{u^{2}}{u_{+}^{2}}$, and the ansatz for the solution, $\phi(\zeta) = (1 - \zeta)^{\alpha} \varphi(\zeta)$, 
lead to the Gauss' hypergeometric differential equation \cite{Birmingham:2001hc}:
\begin{eqnarray}
\zeta (1 - \zeta) \frac{d^2 \varphi(\zeta)}{d \zeta^2} + \left[ c - (1 + a + b)\,\zeta\right] 
\frac{d\,\varphi(\zeta)}{d \zeta} - a\,b\, \varphi(\zeta) = 0 ,
\end{eqnarray}
where $a = \alpha + i\,\tilde{h}$, $b = \alpha - i\,\tilde{h}$, and $c = 1$ with $\tilde{h} = h\, u_{+}$. 
The parameter $\alpha$ is determined by the mass parameter of the scalar field,
\begin{eqnarray}
\alpha = \Delta_{\pm} = \frac{1 \pm \sqrt{1 + m^2}}{2} .
\end{eqnarray}
We thus obtain the solution using the Gauss' hypergeometric function $F(a, b, c\,; \zeta)$ \cite{Bateman1953}:
\begin{eqnarray} 
\phi_{\textrm{sol}}(\zeta) = C\, (1 - \zeta)^{\Delta}\, F (\Delta + i\,\tilde{h}, \Delta - i\,\tilde{h}, 1\,; \zeta) 
\quad \left( \zeta = 1 - \frac{u^2}{u_{+}^{2}} \right) ,
\label{eq: scalar soln with magnetic field}
\end{eqnarray}
where $C$ is a normalization constant and we choose $\alpha = \Delta_{-} \equiv \Delta$ without loss of generality.\footnote{Using the identity of the Gauss' 
hypergeometric function, we can show $(1-\zeta)^{\Delta_{+}} F(\Delta_{+}+i \tilde{h}, \Delta_{+}-i \tilde{h}, 1\,; \zeta) = 
(1-\zeta)^{\Delta_{-}} F(\Delta_{-}+i \tilde{h}, \Delta_{-} - i \tilde{h}, 1\,; \zeta)$ using the relation 
$\Delta_{+} = 1 - \Delta_{-}$.}  
Actually, we have another linearly independent solution which is logarithmically 
divergent at the horizon ($\zeta = 0$), and thus we discard this solution for the regularity inside the bulk.  
We should notice that, for $\Delta < 0$~ (or $m^2 > 0$), this solution diverges at $u = 0$ and 
the position of the boundary should be slightly 
shifted to $u = \varepsilon$ for well-defined calculations, where $\varepsilon$ is a cutoff parameter ($0 < \varepsilon \ll 1$).   
Following the holographic scheme \cite{Freedman:1998tz, Klebanov:1999tb}, 
we take the radial solution normalized 
at the shifted boundary at $u = \varepsilon$, 
\begin{eqnarray}
\phi_{\textrm{sol}}(u) = \left(\frac{u^{2}}{\varepsilon^{2}}\right)^{\Delta}\,  
\frac{F\left(\Delta + i\,\tilde{h}, \Delta - i\,\tilde{h}, 1\,; 
1-u^{2}/u_{+}^{2}\right)}{F\left(\Delta + i\,\tilde{h}, \Delta - i\,\tilde{h}, 1\,; 1-\varepsilon^{2}/u_{+}^{2}\right)} ,
\end{eqnarray} 
and fix the normalization constant,  
\begin{eqnarray}
C \simeq \left(\frac{u_{+}^{2}}{\varepsilon^{2}}\right)^{\Delta}\left(\frac{\Gamma(1 - \Delta - i\,\tilde{h}) 
\Gamma(1 - \Delta + i\,\tilde{h})}{\Gamma( 1 - 2 \Delta )}\right) , 
\label{eq: normalization constant with h}
\end{eqnarray}
where we used the asymptotic form of the hypergeometric function near $\zeta=1$. 

In order to obtain the on-shell action, $S_{\textrm{hol}}\left[\phi_{\textrm{sol}}\right]$, the partial integration 
is performed in the kinetic term in the scalar action (\ref{eq: bulk scalar action}):
\begin{eqnarray}
S_{\textrm{sc}} = - \int\!d^{3}x \sqrt{g}~ \phi^{*} \left[\frac{1}{\sqrt{g}}\, D_{\mu} 
\left(\sqrt{g}\,g^{\mu\nu} D_{\nu} \phi \right) - m^2 \phi\right] + 
\int_{u=\varepsilon}\!\!\!\!\! d^{2}x~ \bigl[\sqrt{g}\,g^{uu} \phi^{*} \partial_{u} \phi \bigr] .
\end{eqnarray}  
The bulk term vanishes upon the solution of the scalar equation of motion 
(\ref{eq: scalar equation}), and thus we only need to evaluate the boundary term at
 $u = \varepsilon$.\footnote{Other two coordinates ($t , \theta$) are periodic and have no boundary.}  
Using the analytic continuation formula for the hypergeometric function \cite{Bateman1953},  
we have the asymptotic expansion of $\phi_{\textrm{sol}}(u)$ near $u=0$\,:
\begin{eqnarray} 
\phi_{\textrm{sol}}(u) \simeq \gamma \left[ \left(\frac{u^{2 \Delta}}{u_{+}^{2 \Delta}}\right) + 
 \left(\frac{\Delta^2 + \tilde{h}^2}{2\,\Delta}\right) \left(\frac{u^{2 \Delta + 2}}{u_{+}^{2 \Delta + 2}}\right) \right] 
+ \xi \left[\left(\frac{u^{2 - 2 \Delta}}{u_{+}^{2 - 2 \Delta}}\right) + 
\left(\frac{(1 - \Delta)^2 + \tilde{h}^2}{2\,(1 - \Delta)}\right)\left(\frac{u^{4 - 2 \Delta}}{u_{+}^{4 - 2 \Delta}}\right) \right].  \nonumber \\
\end{eqnarray}
with the coefficients $\gamma = C\,\frac{\Gamma(1 - 2\,\Delta)}{\Gamma\left(1 - \Delta - i\,\tilde{h}\right) \Gamma\left(1 - \Delta + i\,\tilde{h}\right)}$ and $\xi = 
C\,\frac{\Gamma(2\,\Delta - 1)}{\Gamma(\Delta + i\,\tilde{h}) \Gamma(\Delta - i\,\tilde{h})}$. 
From this asymptotic expansion, the leading terms in the boundary term of the on-shell action near the boundary is given by
\begin{eqnarray}
S_{\textrm{sc}} \simeq \phi_{0}^{2} \int\!d^{2}x\left[ 2\,\gamma^{2} \Delta 
\left(\frac{\varepsilon^{-2 + 4 \Delta}}{u_{+}^{4 \Delta}}\right) 
+\, \gamma^{2} \left(2\,(\Delta^{2} + \tilde{h}^{2}) + \frac{\tilde{h}^{2} - \Delta^{2}}{\Delta}\right) \left(\frac{\varepsilon^{4 \Delta}}{u_{+}^{2 + 4 \Delta}}\right) + 
\left(\frac{2\,\gamma\,\xi}{u_{+}^{2}} \right) \right].~  
\label{eq: scalar boundary term}
\end{eqnarray} 
Note that the boundary action has the divergent terms  
in the limit of $\varepsilon \rightarrow 0$, and 
thus we should subtract such terms using  
the holographic renormalization method  
\cite{Skenderis:2002wp, Fukuma:2002sb}.
Following the recipe of holographic renormalization, 
we add the local counter term at the boundary ($u = \varepsilon$) which cancels the divergent terms,  
\begin{eqnarray}
S_{\textrm{ct}} = - \left[2\,\Delta + \left(\frac{\tilde{h}^{2}}{\Delta}\right)\left(\frac{\varepsilon^{2}}{u_{+}^{2}}\right)\right]\, 
\int_{u = \varepsilon}\!\!\!\!\!d^{2}x~ \sqrt{h}~ \left|\phi(\varepsilon, x)\right|^{2} ,
\end{eqnarray}
and obtain the renormalized on-shell action of the bulk scalar field:
\begin{eqnarray}
S_{\textrm{ren}} = S_{\textrm{sc}} + S_{\textrm{ct}} =  
\left(\frac{2\, \phi_{0}^{2}\, \gamma\, \xi \left(1 - 2 \Delta\right)}{u_{+}^{2}} \right) 
+\, {\cal O}\left(\varepsilon^{2 + 4 \Delta}\right) . 
\end{eqnarray}
where we have assumed $- 1/2 < \Delta < 0$.  
Using the scaling dimension $d = 1 + \sqrt{1+m^{2}} = 2\,(1 - \Delta)$,  
we obtain the free energy of the XXZ spin chain from $S_{\textrm{ren}} = F_{\textrm{xxz}}/T$:
\begin{eqnarray}
F_{\textrm{xxz}} = 4\, \phi_{0}^{2} \left(d-1\right)^2 \varepsilon^{2 d - 4}\, 
(2 \pi\, T_{BH})^{2 d - 2} \sin \pi d  
\left(\frac{\Gamma(\frac{d}{2} - i\,\tilde{h})\,\Gamma(\frac{d}{2} + i\,\tilde{h})}{\Gamma(1 - \frac{d}{2} + i\,\tilde{h})\, 
\Gamma(1 - \frac{d}{2} - i\,\tilde{h})}\right) \Gamma(1 - d)^{2} ,  
\label{eq: holographic free energy}
\end{eqnarray}
where the explicit forms of constants, $\gamma$, $\xi$ and $C$, have been recovered and 
the expression of the Hawking temperature, $u_{+} = \frac{1}{2 \pi\, T_{BH}}$, has been used.  
Under the parameter matching between the holographic dual theory and the XXZ spin chain 
(with the lattice spacing $a$) summarized in Table \ref{tab: parameter matching},  
the resulting free energy (\ref{eq: holographic free energy}), together with the CS results 
(\ref{eq: zero temperature free energy}) and (\ref{eq: finite size scaling free energy}),  completely reproduces the free energy of the XXZ spin chain 
(with $\frac{2}{3} < \beta^{2} < 1$), 
which is calculated by the field theory and integrability techniques  \cite{Lukyanov:1998}, 
including the numerical factors.\footnote{Note that the definition of external magnetic field is different between 
ours and the reference \cite{Lukyanov:1998} by the factor of 2.}
\begin{table}[h]
  \centering
  \begin{tabular}{|lc|c|}
    \hline
    {}\hspace*{0.25cm}Parameter in & \hspace*{-1.5cm}dual theory & matching with spin chain\\
    \hline
    cut off   & $\varepsilon$  & $a$ \\
    CS coupling   & $k$         & $1/\beta^2$ \\
    charge    & $q$         & $1/\beta^2$ \\
    mass      & $m^2$       & $4(1-\beta^2)/\beta^{4}$ \\
    perturbation strength  & $\phi_{0}$  & $\lambda(\beta)/4(2\beta^{-2}-1)$ \\
    \hline
    \end{tabular}
    \caption{Parameter matching: XXZ spin chain anisotropy is $\delta = -\cos(\pi\beta^2)$. 
    Irrelevant coupling constant $\lambda(\beta)$ is given by Eq.\,(2.24) in \cite{Lukyanov:1998}.}
    \label{tab: parameter matching}
\end{table}

It should be noticed that, in the massless limit of the bulk scalar field, which corresponds to 
the isotropic limit ($\delta = 1$) of the spin chain,  
the scalar solution (\ref{eq: scalar soln with magnetic field}) has the logarithmically divergent terms 
in the asymptotic expansion at $u \sim 0$, whose coefficients involve the digamma function, 
and the logarithmic corrections to thermodynamic quantities of 
the isotropic spin chain \cite{Affleck:1994, Lukyanov:1998} can be reproduced.   

\section{Discussion}
We have constructed the holographic dual theory of the XXZ quantum spin chain based on the CS gauge theory 
with a charged scalar field,\footnote{A similar setup has been also discussed for the holographic dual 
description of the Kondo effect \cite{Erdmenger:2013dpa, Erdmenger:2015spo, Erdmenger:2015xpq}.} and calculated the thermodynamic quantities using the holographic techniques, 
which are completely consistent with the exact results of the spin chain. 
This is, to our best knowledge, the first example which has the precise matching of the physical quantities 
(including the numerical factors) between the holographic dual theory and 
the quantum many-body system in condensed matters, and 
this precise duality can lead to not only new insights from higher dimensional perspectives 
but also new analytical methods in quantum many-body systems.  
The nontrivial matching between the bulk scalar action and the free energy of the spin chain 
can give a physical background to the ODE/IM correspondence \cite{Dorey:2007zx}, which is 
the correspondence between ordinary differential equations and quantum integrable models, 
based on the black holes through the holographic duality.    
Furthermore the holographic dual theory can have the applications 
to the dynamics of one-dimensional quantum many-body systems 
in real materials, 
such as various transports in Tomonaga-Luttinger liquids \cite{Fradkin}, 
quantum spin chain \cite{HirobeSpinon}, 
and spacetime-emergent materials \cite{Hashimoto:2022aso}. 

\vspace*{0.4cm}    
\noindent
{\large {\bf Acknowledgement}}

\noindent
This work was partially supported by ERATO (No. JPMJER1402), CREST 
(Nos. JPMJCR20C1, JPMJCR20T2) 
from JST, Japan; Grant-in-Aid for 
Scientific Research (S) (No. JP19H05600), Grant-in-Aid for 
Transformative Research Areas (No. JP22H05114) from JSPS KAKENHI, Japan. 
This work was also supported by Institute of AI and Beyond of the
University of Tokyo, Japan.

\end{document}